\documentclass[acmlarge]{acmart}

\AtBeginDocument{%
  }

\setcopyright{none}

\acmJournal{ONLINE}

\begin{document}

\title{Quantifying Community Evolution in Developer Social Networks: Proof of Indices' Properties}

\author{Liang Wang}
\email{wl@nju.edu.cn}
\orcid{0000-0001-5444-748X}
\affiliation{%
  \institution{State Key Laboratory for Novel Software Technology, Nanjing University}
  \streetaddress{163 Xianlin Ave.}
  \city{Nanjing}
  \country{China}
}

\author{Ying Li}
\email{mg21070006@smail.nju.edu.cn}
\orcid{0000-0002-4637-1742}
\affiliation{%
  \institution{State Key Laboratory for Novel Software Technology, Nanjing University}
  \streetaddress{163 Xianlin Ave.}
  \city{Nanjing}
  \country{China}
}

\author{Jierui Zhang}
\email{jieruizhang@smail.nju.edu.cn}
\orcid{0000-0002-7290-790X}
\affiliation{%
  \institution{State Key Laboratory for Novel Software Technology, Nanjing University}
  \streetaddress{163 Xianlin Ave.}
  \city{Nanjing}
  \country{China}
}

\author{Xianping Tao}
\email{txp@nju.edu.cn}
\orcid{0000-0002-5536-3891}
\affiliation{%
  \institution{State Key Laboratory for Novel Software Technology, Nanjing University}
  \streetaddress{163 Xianlin Ave.}
  \city{Nanjing}
  \country{China}
}

\renewcommand{\shortauthors}{Liang Wang, Ying Li, Jierui Zhang, and Xianping Tao}

\begin{abstract}
The document provides the proof to properties of community evolution indices including community \textit{split} and \textit{shrink} in paper: Liang Wang, Ying Li, Jierui Zhang, and Xianping Tao. 2022. QuantifyingCommunity Evolution in Developer Social Networks. InProceedings of the30th ACM Joint European Software Engineering Conference and Symposiumon the Foundations of Software Engineering (ESEC/FSE ’22), November 14–18, 2022, Singapore, Singapore.ACM, New York, NY, USA, 12 pages. https://doi.org/10.1145/3540250.3549106.
Proof to properties of community \textit{merge} and \textit{expand} is similar.
\end{abstract}



\maketitle

\section{Brief Introduction to the Properties of Community Split and Shrink Indices}

Let $\mathcal{I}^{\psi}_{c_{t,i}}$ and $\mathcal{I}^{\eta}_{c_{t,i}}$ denote the community \textit{split} and \textit{shrink} indices, respectively.
Without loss of generality, we assume $m\geq 1$. 
The properties of the two indices are as follows.

\textbf{P-1.}
$\mathcal{I}^{\psi}_{c_{t,i}}$ and $\mathcal{I}^{\eta}_{c_{t,i}}$ are \underline{strictly monotonic increasing} functions of $m$, given $0 < \eta_i < 1$, and $\hat{\psi}_{i,j} = \frac{1}{m}, j=1,2,\cdots,m$.

\textbf{P-2.}
$\mathcal{I}^{\psi}_{c_{t,i}}$ / $\mathcal{I}^{\eta}_{c_{t,i}}$ is a \underline{strictly monotonic decreasing / increasing} function of $\eta_i$, respectively, for $\eta_i > 0$, given $m>1$, and member migration distribution $\hat{\psi}_{i,j}, j=1,2,\cdots,m$ with $\mathcal{H}_{c_{t,i}} > 0$.

\textbf{P-3.}
Given $m$ and $\eta_i$, the \underline{maximum split index} $\mathcal{I}^{\psi}_{c_{t,i}} = (1-\eta_i)\mathcal{H}^*_{t\rightarrow t+1}$ is obtained when the members of $c_{t,i}$ migrate to the communities detected in the next step with a even distribution, i.e., when we have $\hat{\psi}_{i,j} = \frac{1}{m}, j=1,2,\cdots,m$.
And the \underline{minimum split index} $\mathcal{I}^{\psi}_{c_{t,i}} = 0$ is obtained when $m=1$ or all the members of  $c_{t,i}$ who stay in the project migrate to a single community in the next step, i.e., there exists a $j'$-th community in time $t+1$ that $\hat{\psi}_{i,j'} = 1$ and $\hat{\psi}_{i,j} = 0, \forall j \neq j'$, resulting in $\mathcal{H}_{c_{t,i}}=0$ and $\mathcal{I}^{\psi}_{c_{t,i}} = 0$.
    
\textbf{P-4.}
Given $m>1$ and $\eta_i$, the \underline{maximum shrink index} $\mathcal{I}^{\eta}_{c_{t,i}} = \eta_i\mathcal{H}^*_{t\rightarrow t+1}$ is obtained when the corresponding split index is minimized, i.e., all stayed members of community $c_{t,i}$ migrate to a single community in the next step.
And the \underline{minimum shrink index} $\mathcal{I}^{\eta}_{c_{t,i}} = \eta_i^2\mathcal{H}^*_{t\rightarrow t+1}$ is obtained when the members of $c_{t,i}$ migrate evenly to communities in time $t+1$. For the special case of $m=1$, the shrink index is only determined by $\eta_i$.

\begin{figure}[h]
    \centering
    \includegraphics[width = 0.7\linewidth]{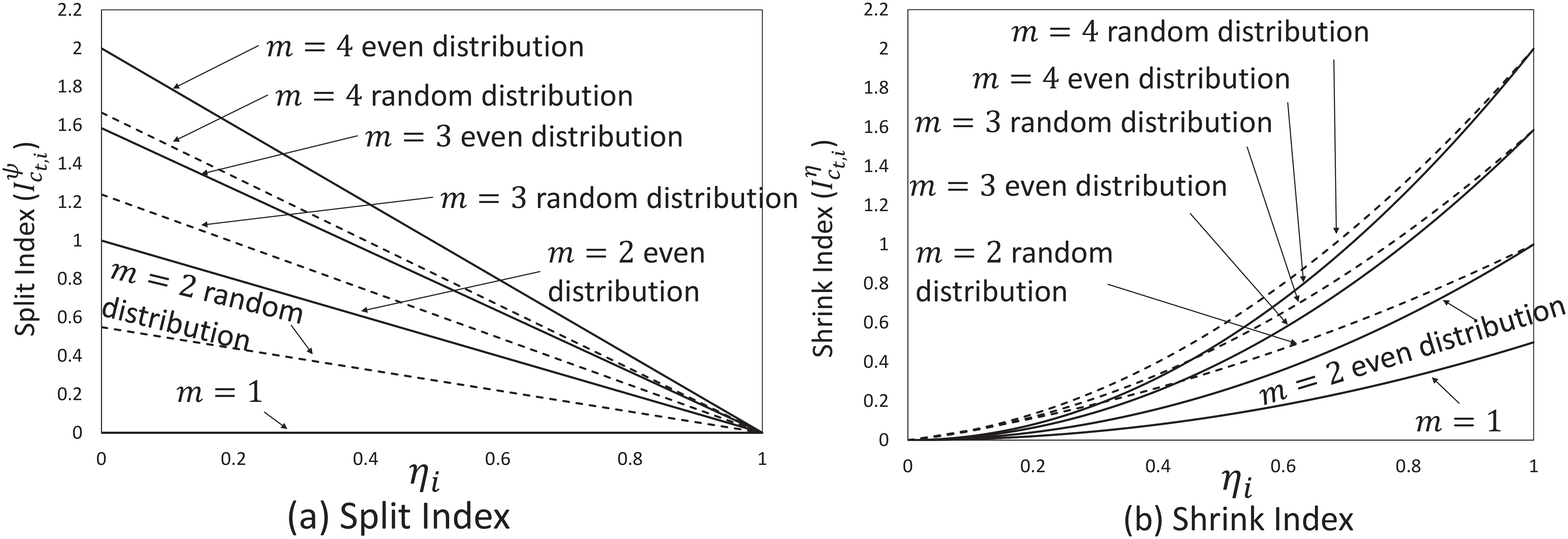}
    \caption{Curves of the \textit{split} and \textit{shrink} indices under different conditions specified by $m$, $\eta_i$, and $\hat{\psi}_{i,j}$. The even distribution corresponds to case $\hat{\psi}_{i,j} = \frac{1}{m},  j=1,\cdots,m$. The random distribution is obtained by randomly assigning  $\hat{\psi}_{i,j}$'s values.
    }
    \label{fig:split_shrink_simulate}
\end{figure}

Fig. \ref{fig:split_shrink_simulate} illustrates the curves of the split and shrink indices under different conditions, from which we can find correspondence to the above properties.

\section{Proof of The Properties}

The proof of the above four properties are as follows.

\noindent\textbf{P-1.} $\mathcal{I}^{\psi}_{c_{t,i}}$ and $\mathcal{I}^{\eta}_{c_{t,i}}$ are \underline{strictly monotonic increasing} functions of $m$, given $0 < \eta_i < 1$, and $\hat{\psi}_{i,j} = \frac{1}{m}, j=1,2,\cdots,m$.

\noindent\textbf{Proof.} 
First, for community split index $\mathcal{I}^{\psi}_{c_{t,i}}$, take the partial derivative of $\mathcal{I}^{\psi}_{c_{t,i}}$ with respect to $m$ gives us:
\begin{equation}\label{eq:par_split_m}
    \frac{\partial \mathcal{I}^{\psi}_{c_{t,i}}}{\partial m} = \frac{\partial (1-\eta_{i}) \mathcal{H}_{c_{t,i}}}{\partial m} = (1-\eta_{i})\frac{\partial \mathcal{H}_{c_{t,i}}}{\partial m}.
\end{equation}

When $\hat{\psi}_{i,j} = \frac{1}{m}, j=1,2,\cdots,m$, we have $\mathcal{H}_{c_{t,i}} = -\sum_{j=1}^m \hat{\psi}_{i,j} \log_2(\hat{\psi}_{i,j}) = -\log_2 \frac{1}{m}$.
As a result, we have
\begin{equation}\label{eq:par_h_m}
    \frac{\partial \mathcal{H}_{c_{t,i}}}{\partial m} = \frac{1}{m \ln(2)}.
\end{equation}

Substitute Eq. (\ref{eq:par_h_m}) into Eq. (\ref{eq:par_split_m}), we have
\begin{equation}\label{eq:par_split_m_final}
    \frac{\partial \mathcal{I}^{\psi}_{c_{t,i}}}{\partial m} = \frac{1-\eta_{i}}{m\ln(2)}.
\end{equation}

We can see from Eq. (\ref{eq:par_split_m_final}) that $\frac{\partial \mathcal{I}^{\psi}_{c_{t,i}}}{\partial m}>0$, for $m \geq 1$, and $0 \leq \eta_i < 1$. 

As a result, the community split index $\mathcal{I}^{\psi}_{c_{t,i}}$ is a \underline{strictly monotonic increasing} function of $m$, for $m \geq 1$, $0 \leq \eta_i < 1$, and $\hat{\psi}_{i,j} = \frac{1}{m}, j=1,2,\cdots,m$.

For the case of $\eta_i = 1$, we always have $\mathcal{I}^{\psi}_{c_{t,i}} = 0, \forall m\geq 1$, and regardless the distribution of $\hat{\psi}_{i,j}, j=1,2,\cdots,m$, which is intuitively correct that a community cannot split if all of its members leaves the project.

\vspace{1cm}

Next, for community shrink index $\mathcal{I}^{\eta}_{c_{t,i}}$, we also take its partial derivative with respect to $m$, which is
\begin{equation}\label{eq:par_shrink_m}
    \frac{\partial \mathcal{I}^{\eta}_{c_{t,i}}}{\partial m} = \frac{\partial \eta_{i}(\mathcal{H}^*_{t\rightarrow t+1} - \mathcal{I}^{\psi}_{c_{t,i}} + \sigma_{\eta_i})}{\partial m} = \eta_{i}(\frac{\partial \mathcal{H}^*_{t\rightarrow t+1}}{\partial m} - \frac{\partial \mathcal{I}^{\psi}_{c_{t,i}}}{\partial m}).
\end{equation}

Since $\frac{\partial \mathcal{H}^*_{t\rightarrow t+1}}{\partial m} = \frac{\partial -\log_2(\frac{1}{m})}{\partial m} = \frac{1}{m \ln(2)}$, and $\frac{\partial \mathcal{I}^{\psi}_{c_{t,i}}}{\partial m} = \frac{1-\eta_{i}}{m\ln(2)}$ following Eq. (\ref{eq:par_split_m_final}), we have 
\begin{equation}\label{eq:par_shrink_m_final}
    \frac{\partial \mathcal{I}^{\eta}_{c_{t,i}}}{\partial m} = \frac{\eta_{i}^2}{m\ln(2)}.
\end{equation}

From Eq. (\ref{eq:par_shrink_m_final}), we have $\frac{\partial \mathcal{I}^{\eta}_{c_{t,i}}}{\partial m} > 0$ for $m \geq 1$, $0 < \eta_i \leq 1$, and $\hat{\psi}_{i,j} = \frac{1}{m}, j=1,2,\cdots,m$.

For the case of $\eta_i = 0$, we always have $\mathcal{I}^{\eta}_{c_{t,i}} = 0, \forall m\geq 1$, given any member migration distribution $\hat{\psi}_{i,j}, j=1,2,\cdots,m$, which is reasonable because a community does not shrink if none of its members leave the project, regardless the number of communities detected in the next step (i.e., $m$).

\vspace{1cm}

Combining the above two results, we show that: $\mathcal{I}^{\psi}_{c_{t,i}}$ and $\mathcal{I}^{\eta}_{c_{t,i}}$ are \underline{strictly monotonic increasing} functions of $m$, given $0 < \eta_i < 1$, and $\hat{\psi}_{i,j} = \frac{1}{m}, j=1,2,\cdots,m$.

$\hfill\Box$

\vspace{1cm}

\noindent\textbf{P-2.} 
$\mathcal{I}^{\psi}_{c_{t,i}}$ / $\mathcal{I}^{\eta}_{c_{t,i}}$ is a \underline{strictly monotonic decreasing / increasing} function of $\eta_i$, respectively, for $\eta_i > 0$, given $m>1$, and member migration distribution $\hat{\psi}_{i,j}, j=1,2,\cdots,m$ with $\mathcal{H}_{c_{t,i}} > 0$.

\noindent\textbf{Proof.}
First, we show that community split index $\mathcal{I}^{\psi}_{c_{t,i}}$ is a \underline{strictly monotonic decreasing} function of $\eta_i$ under the given conditions.
The partial derivative of $\mathcal{I}^{\psi}_{c_{t,i}}$ with respect to $\eta_i$ is
\begin{equation}\label{eq:par_split_eta}
    \frac{\partial \mathcal{I}^{\psi}_{c_{t,i}}}{\partial \eta_i} = \frac{\partial  (1-\eta_{i}) \mathcal{H}_{c_{t,i}}}{\partial \eta_i} = -\mathcal{H}_{c_{t,i}}.
\end{equation}

As a result, we have $\frac{\partial \mathcal{I}^{\psi}_{c_{t,i}}}{\partial \eta_i} < 0$, meaning that $\mathcal{I}^{\psi}_{c_{t,i}}$ is a \underline{strictly monotonic decreasing} function of $\eta_i$, as long as $\mathcal{H}_{c_{t,i}} > 0$.

We have $\mathcal{H}_{c_{t,i}}=0$ and thus $\mathcal{I}^{\psi}_{c_{t,i}} = 0, \forall \eta_i \in [0,1]$ when only one of the $\hat{\psi}_{i,j}$'s is one with rest of the $\hat{\psi}_{i,j}$'s equal to zero (including the case that $m=1$).
Intuitively, this case means that a community is not regarded as splitting if all of its remaining members migrate to a single community in the next step, regardless the amount of members who leave the project.

\vspace{1cm}

Next, we show that community shrink index $\mathcal{I}^{\eta}_{c_{t,i}}$ is a \underline{strictly monotonic increasing} function of $\eta_i$ under the given conditions.
Taking the partial derivative of $\mathcal{I}^{\eta}_{c_{t,i}}$ with respect to $\eta_i$ gives us
\begin{equation}\label{eq:par_shrink_eta}
\begin{split}
    \frac{\partial \mathcal{I}^{\eta}_{c_{t,i}}}{\partial \eta_i} & = \frac{\partial \eta_{i}(\mathcal{H}^*_{t\rightarrow t+1} - \mathcal{I}^{\psi}_{c_{t,i}} + \sigma_{\eta_i})}{\partial \eta_i} = (\mathcal{H}^*_{t\rightarrow t+1} - \mathcal{I}^{\psi}_{c_{t,i}} + \sigma_{\eta_i}) + \eta_{i}\frac{\partial (\mathcal{H}^*_{t\rightarrow t+1} - \mathcal{I}^{\psi}_{c_{t,i}} + \sigma_{\eta_i})}{\partial \eta_i}\\
    & = \mathcal{H}^*_{t\rightarrow t+1} - (1-\eta_i)\mathcal{H}_{c_{t,i}} + \sigma_{\eta_i} + \eta_i \frac{\partial \mathcal{I}^{\psi}_{c_{t,i}}}{\partial \eta_i} + \eta_i \frac{\partial \sigma_{\eta_i}}{\partial \eta_i}.
\end{split}
\end{equation}

The forth term in Eq. (\ref{eq:par_shrink_eta}) is $\eta_i\frac{\partial \mathcal{I}^{\psi}_{c_{t,i}}}{\partial \eta_i} = \eta_i\frac{\partial (1-\eta_i)\mathcal{H}_{c_{t,i}}}{\partial \eta_i} = -\eta_i\mathcal{H}_{c_{t,i}}$.
And the last term in  Eq. (\ref{eq:par_shrink_eta}) is $\eta_i \frac{\partial \sigma_{\eta_i}}{\partial \eta_i} = 0.5\eta_i$ when $m=1$, and  $\eta_i \frac{\partial \sigma_{\eta_i}}{\partial \eta_i} = 0$ when $m>1$ following the definition of $\sigma_{\eta_i}$.

For the case of $m>1$, Eq. (\ref{eq:par_shrink_eta}) can be written as
\begin{equation}\label{eq:par_shrink_eta_final}
    \frac{\partial \mathcal{I}^{\eta}_{c_{t,i}}}{\partial \eta_i} = \mathcal{H}^*_{t\rightarrow t+1} - (1-2\eta_i)\mathcal{H}_{c_{t,i}}.
\end{equation}

Because $\mathcal{H}^*_{t\rightarrow t+1}$ is the maximum entropy, we have $\mathcal{H}^*_{t\rightarrow t+1} > 0$, $\mathcal{H}^*_{t\rightarrow t+1} \geq \mathcal{H}_{c_{t,i}} \geq 0$, and $(1-2\eta_i) < 1$ when $\eta_i>0$.
We then have $\frac{\partial \mathcal{I}^{\eta}_{c_{t,i}}}{\partial \eta_i} > 0$ for $\eta_i>0$ and $m>1$.

For the case of $m=1$, we have $\sigma_{\eta_i} = \eta_i \frac{\partial \sigma_{\eta_i}}{\partial \eta_i} = 0.5\eta_i$, and $\mathcal{H}^*_{t\rightarrow t+1} = \mathcal{H}_{c_{t,i}} = 0$.
Eq. (\ref{eq:par_shrink_eta}) then becomes
\begin{equation}\label{eq:par_shrink_eta_final_0}
    \frac{\partial \mathcal{I}^{\eta}_{c_{t,i}}}{\partial \eta_i} = \eta_i.
\end{equation}

We also have $\frac{\partial \mathcal{I}^{\eta}_{c_{t,i}}}{\partial \eta_i} > 0$ for $\eta_i>0$ and $m=1$.
As a result, $\mathcal{I}^{\eta}_{c_{t,i}}$ is a \underline{strictly monotonic increasing} function of $\eta_i$ for $\eta_i > 0$, and $m \geq 1$.

\vspace{1cm}

Summarizing the above, we show that: $\mathcal{I}^{\psi}_{c_{t,i}}$ / $\mathcal{I}^{\eta}_{c_{t,i}}$ is a \underline{strictly monotonic decreasing / increasing} function of $\eta_i$, respectively, for $\eta_i > 0$, given $m>1$, and the distribution of member migration $\hat{\psi}_{i,j}, j=1,2,\cdots,m$ with $\mathcal{H}_{c_{t,i}} > 0$.

$\hfill\Box$

\vspace{1cm}

\noindent\textbf{P-3.} Given $m$ and $\eta_i$, the \underline{maximum split index} $\mathcal{I}^{\psi}_{c_{t,i}} = (1-\eta_i)\mathcal{H}^*_{t\rightarrow t+1}$ is obtained when the members of $c_{t,i}$ migrate to the communities detected in the next step with a even distribution, i.e., when we have $\hat{\psi}_{i,j} = \frac{1}{m}, j=1,2,\cdots,m$.
And the \underline{minimum split index} $\mathcal{I}^{\psi}_{c_{t,i}} = 0$ is obtained when $m=1$ or all the members of  $c_{t,i}$ who stay in the project migrate to a single community in the next step, i.e., there exists a $j'$-th community in time $t+1$ that $\hat{\psi}_{i,j'} = 1$ and $\hat{\psi}_{i,j} = 0, \forall j \neq j'$, resulting in $\mathcal{H}_{c_{t,i}}=0$ and $\mathcal{I}^{\psi}_{c_{t,i}} = 0$.

\noindent\textbf{Proof.}
$\mathcal{I}^{\psi}_{c_{t,i}} = (1-\eta_{i}) \mathcal{H}_{c_{t,i}}$ is a monotonic increasing function of $\mathcal{H}_{c_{t,i}}$ for $0\leq \eta_i \leq 1$.
Referring to the properties of information entropy \cite{shannon1948mathematical}, entropy $\mathcal{H}_{c_{t,i}} = \mathcal{H}^*_{t\rightarrow t+1} = -\log_2 \frac{1}{m}$ is a maximum when $\hat{\psi}_{i,j} = \frac{1}{m}, j=1,2,\cdots,m$.
And the minimum value of $\mathcal{H}_{c_{t,i}} = 0$ is obtained when only one of the $\hat{\psi}_{i,j}$'s equals to one and others equal to zero, which also includes the case of $m=1$.
As a result, the maximum split index is $\mathcal{I}^{\psi}_{c_{t,i}} = (1-\eta_i)\mathcal{H}^*_{t\rightarrow t+1} = -(1-\eta_i)\log_2 \frac{1}{m}$ given $\eta_i$.
When $\eta_i = 0$, we have $\mathcal{I}^{\psi}_{c_{t,i}} = -\log_2 \frac{1}{m}$ being the maximum possible value for the split index, which is only determined by $m$.
And the minimum split index is  $\mathcal{I}^{\psi}_{c_{t,i}} = 0$.

$\hfill\Box$

\vspace{1cm}

\noindent\textbf{P-4.} Given $m>1$ and $\eta_i$, the \underline{maximum shrink index} $\mathcal{I}^{\eta}_{c_{t,i}} = \eta_i\mathcal{H}^*_{t\rightarrow t+1}$ is obtained when the corresponding split index is minimized, i.e., all stayed members of community $c_{t,i}$ migrate to a single community in the next step.
And the \underline{minimum shrink index} $\mathcal{I}^{\eta}_{c_{t,i}} = \eta_i^2\mathcal{H}^*_{t\rightarrow t+1}$ is obtained when the members of $c_{t,i}$ migrate evenly to communities in time $t+1$. For the special case of $m=1$, the shrink index is only determined by $\eta_i$.

\noindent\textbf{Proof.}
Given $m>1$ and $\eta_i$, the community shrink index $\mathcal{I}^{\eta}_{c_{t,i}} = \eta_{i}(\mathcal{H}^*_{t\rightarrow t+1} - \mathcal{I}^{\psi}_{c_{t,i}} + \sigma_{\eta_i})$ is a monotonic decreasing function of the community split index $\mathcal{I}^{\psi}_{c_{t,i}}$.
Referring to Property 3 presented above, we have the shrink index maximized when the split index is minimized, i.e., $\mathcal{H}_{c_{t,i}} = 0$.
The maximum shrink index is given by $\mathcal{I}^{\eta}_{c_{t,i}} = \eta_i\mathcal{H}^*_{t\rightarrow t+1} = -\eta_i\log_2\frac{1}{m}$.
And the minimum shrink index $\mathcal{I}^{\eta}_{c_{t,i}} = \eta_i^2\mathcal{H}^*_{t\rightarrow t+1} =  -\eta_i^2\log_2\frac{1}{m}$ when $\mathcal{I}^{\psi}_{c_{t,i}} = (1-\eta_i)\mathcal{H}^*_{t\rightarrow t+1}$.

In the above analysis we have $\sigma_{\eta_i} = 0$ because we assume $m>1$.
When $m=1$, we have $\sigma_{\eta_i} = 0.5\eta_i$, and $\mathcal{H}^*_{t\rightarrow t+1} = \mathcal{H}_{c_{t,i}} = 0$.
The shrink index is $\mathcal{I}^{\eta}_{c_{t,i}} = 0.5\eta_i^2$, which is only determined by $\eta_i$.

Following the above analysis, if we consider the change of $\eta_i$, we can find that the maximum and minimum possible value of the shrink index is $\mathcal{I}^{\eta}_{c_{t,i}} = -\log_2 \frac{1}{m}$ when $\eta_i=1$, and $\mathcal{I}^{\eta}_{c_{t,i}} = 0$ when $\eta_i=0$, respectively.

$\hfill\Box$

\vspace{1cm}

From properties \textbf{P-3} and \textbf{P-4} we can further see that given $m>1$, the community split and shrink indices vary in the same range given by $\mathcal{I}^{\psi}_{c_{t,i}}\in[0, -\log_2\frac{1}{m}]$, and $\mathcal{I}^{\eta}_{c_{t,i}}\in[0, -\log_2\frac{1}{m}]$, with different values of $\eta_i$ and the distribution of member migration specified by $\hat{\psi}_{i,j}, j=1,2,\cdots,m$.
As a result, it is feasible for us to draw meaningful results, such as \textit{the community shows a stronger trend of splitting / shrinking}, by directly comparing the values of community split and shrink indices.

\begin{acks}
We thank all the reviewers for their efforts in improving the paper.
This work is supported by the National Key R\&D Program of China No. 2018AAA0102302, NSFC No. 62172203, Fundamental Research Funds for the Central Universities, and the Collaborative Innovation Center of Novel Software Technology and Industrialization.
\end{acks}

\bibliographystyle{ACM-Reference-Format}
\bibliography{ref}


\begin{thebibliography}{1}


\ifx \showCODEN    \undefined \def \showCODEN     #1{\unskip}     \fi
\ifx \showDOI      \undefined \def \showDOI       #1{#1}\fi
\ifx \showISBNx    \undefined \def \showISBNx     #1{\unskip}     \fi
\ifx \showISBNxiii \undefined \def \showISBNxiii  #1{\unskip}     \fi
\ifx \showISSN     \undefined \def \showISSN      #1{\unskip}     \fi
\ifx \showLCCN     \undefined \def \showLCCN      #1{\unskip}     \fi
\ifx \shownote     \undefined \def \shownote      #1{#1}          \fi
\ifx \showarticletitle \undefined \def \showarticletitle #1{#1}   \fi
\ifx \showURL      \undefined \def \showURL       {\relax}        \fi
\providecommand\bibfield[2]{#2}
\providecommand\bibinfo[2]{#2}
\providecommand\natexlab[1]{#1}
\providecommand\showeprint[2][]{arXiv:#2}

\bibitem[Shannon(1948)]%
        {shannon1948mathematical}
\bibfield{author}{\bibinfo{person}{Claude~Elwood Shannon}.}
  \bibinfo{year}{1948}\natexlab{}.
\newblock \showarticletitle{A mathematical theory of communication}.
\newblock \bibinfo{journal}{\emph{The Bell system technical journal}}
  \bibinfo{volume}{27}, \bibinfo{number}{3} (\bibinfo{year}{1948}),
  \bibinfo{pages}{379--423}.
\newblock


\end{thebibliography}

\end{document}